\documentclass[prb,twocolumn,showpacs,superscriptaddress,floatfix]{revtex4}
\usepackage{graphicx}
\usepackage{dcolumn}
\usepackage{bm}
\begin{document}

\title{Pressure--induced amorphization, crystal--crystal transformations
and the memory glass effect in interacting particles in two dimensions}
\author{S. Bustingorry}
\affiliation{Centro At\'omico Bariloche and Instituto Balseiro,
Comisi\'on Nacional de Energ\'{\i}a At\'omica,
(8400) Bariloche, Argentina}
\author{E. A. Jagla}
\affiliation{The Abdus Salam ICTP, Strada Costiera 11, (34014) Trieste, Italy}
\date{\today}

\begin{abstract}
We study a model of interacting particles in two dimensions to address 
the relation between crystal-crystal transformations and
pressure-induced amorphization. On increasing pressure at very low temperature,
our model undergoes a martensitic crystal-crystal transformation.
The characteristics of the resulting polycrystalline structure depend
on defect density, compression rate, and nucleation and growth barriers.
We find two different limiting cases. In one of them the martensite crystals,
once nucleated, grow easily perpendicularly to the invariant interface, and the final structure
contains large crystals of the different martensite variants. Upon decompression almost every
atom returns to its original position, and the original crystal is fully recovered.
In the second limiting case, after nucleation
the growth of martensite crystals is inhibited by energetic barriers. The final morphology in this
case is that of a polycrystal with a very small crystal size.
This may be taken to be amorphous if we have only access (as experimentally
may be the case) to the angularly averaged structure factor.
However, this `X-ray amorphous' material is anisotropic, and this shows up upon decompression,
when it recovers the original crystalline structure with an orientation correlated
with the one it had prior to compression.
The memory effect  of this X-ray amorphous material is a natural
consequence of the memory effect associated to the underlying martensitic
transformation. We suggest that this kind of mechanism is present in many
of the experimental observations of the memory glass effect, in which a
crystal with the original orientation is recovered from an apparently amorphous
sample when pressure is released.
\end{abstract}

\maketitle

\section{Introduction}\label{s:introduction}

Several crystalline materials are known to become amorphous under changes
of the ambient pressure,
\cite{pony92,shar96,rich97} and this may occur under compression
or decompression.
Among the most thoroughly studied cases, due to its
technological or geophysical importance, we find H$_2$O, SiO$_2$, GeO$_2$,
AlPO$_4$ and other isostructural compounds. The occurrence of
pressure induced amorphization (PIA) is characteristic of
(although not limited to)
tetrahedrally coordinated materials.

The simple statement that a material becomes amorphous under
pressure covers a broad range of different experimental
situations and possible theoretical interpretations. The simplest
possibility is a crystalline material that transforms (upon
change of the external pressure) to a completely amorphous
structure when a well defined transition pressure is reached. We
will call this possibility true PIA (TPIA). This strict case of
PIA is by no means the only possible one. A reason for this is
that typically PIA is claimed to occur (on an experimental basis)
when the obtained sample does not display sharp X-ray diffraction
peaks. We will refer to these samples as `X-ray amorphous' material. A
relatively large amount of local order can still be present in an
X-ray amorphous. For instance, a polycrystal with a sufficiently
small crystal size is an X-ray amorphous material. An eventual transition
from a single-crystal sample to this kind of polycrystal should
be considered, on an experimental basis, to be a PIA transition.
This is a second case of PIA that we will call `weak' (WPIA).

Experimentally, WPIA or some variants of it may be the rule more than the exception.
It is seen in many cases that amorphization occurs through a
sequence of transformations in which one or more intermediate crystalline
phases can be identified. For instance, a crystal-crystal pressure induced
transformation  was shown to occur in $\alpha-$quartz between $21$ and $30$ Gpa
\cite{king93, greg00, hain01}. The new crystalline phases, quartz II and a $P2_1/c$
phase, may be intermediate states in the path toward complete amorphization,
occurring at higher pressures. Yet, the final amorphous material
may well be only an X-ray amorphous material.

A common characteristic of both WPIA and TPIA is that the
starting crystalline sample becomes unstable upon pressure. This
instability does not typically disappear if temperature is
reduced. This is an indication that in most, if not all cases,
the mechanisms of PIA can be discussed in the limit of $T=0$. In
this limit, all structural transformations that may eventually
occur are of a mechanical nature, i.e., the displacement of the
particles during the transformation is strictly guided by the
tendency of the system to minimize its mechanical energy. In
addition to being a simplifying assumption, the consideration of
vanishingly small temperature is conceptually important, as in
this case structural transformations are necessarily related to
mechanical instabilities in the system. We notice that although
we will present all our analysis at $T\sim 0$, this is in
principle applicable also to any temperature below the glass
transition temperature $T_g$ of the fluid phase, where diffusion
effects are negligible. We only need to consider in this case
that the parameters of the model may depend on temperature. Since
at $T<T_g$ diffusive processes are almost totally inhibited, any
structural transition (in particular, PIA) occurring in this
temperature range is of a displacive nature. This raises the
question of the relation between PIA and martensitic
transformations.\cite{nish78} In fact, the discussion of the
similarities between PIA and martensitic transformations
is one of the main aims of the present paper.

There is also a long standing controversy about whether PIA is of
a mechanical or thermodynamical
nature,\cite{shar96,mish96,lyap96,badr98,tse99,grom01} and this in
turn is related to the two possible mechanisms of melting at
higher temperatures: mechanical or thermodynamical.\cite{wolf90}
Thermodynamical melting implies the growth of the fluid phase
when it starts to be more stable than the solid phase. This
growth is heterogeneous and occurs in the presence of extended
nucleation centers, such as grain boundaries or even the surface
of the sample. If these defects do not exist (something that can
be achieved in numerical simulations by the use of
single-crystalline samples with periodic boundary conditions),
then the fluid phase is not able to nucleate until the ultimate
mechanical equilibrium conditions are violated. This is known as
Born's mechanical stability criteria, and the melting
mechanism in this limit is called mechanical melting. Mechanical
melting is homogeneous, as it affects the whole sample at the
same time, and occurs very rapidly once the stability limit is
reached. It has been suggested that these two melting mechanisms
can be extended to describe two different amorphization
mechanisms.\cite{wolf90}


There are quite a few numerical models devised to study PIA in
different materials. In some cases, a thorough description of a
particular material is pursued. In this case, the models range
from first principle density functional calculations
\cite{chel90,pomp94} to molecular dynamics simulations with
different pair-potential interactions.\cite{tsun89,bees90,bing94}
Within these models, PIA can be directly studied in concrete
systems such as H$_2$O or SiO$_2$. A more qualitative approach is
obtained by using core-softened potentials,\cite{hemm70,stel72}
which proved to be very useful for the study of water-like
anomalies and polyamorphism in tetrahedrally coordinated
materials.\cite{jagl98,sadr98,jagl99,jagl01,scal01} The aim of
this approach is not to carry out an analysis of the properties
of a specific material, but to find out general trends and
qualitative behaviors that shed light on the properties of a
large family of materials. This is the approach that we will
follow here.

We study here a two-dimensional model of identical particles
interacting through a spherically symmetric two body potential.
We have recently shown that choosing appropriately the
interaction potential, the model presents TPIA upon pressure
release if we start from the stable high pressure
phase.\cite{bust04} Here we analyze in detail the WPIA that
occurs when we increase pressure starting from the stable low
pressure phase. We show that this transformation is of a
martensitic nature, and analyze the nucleation and growth of the
crystallites of the new phase. Depending on details of the
potential, two different extreme possibilities occur. In one of
them, the new phase grows in large crystallites of the (three)
different martensitic variants, and crystallites grow
perpendicularly to the habit line of the martensitic
transformation. This is a case of ideal martensitic
transformation. In the other limiting case, the new crystallites
nucleate but can hardly grow. In this case, a WPIA scenario is
realized. The final sample is X-ray amorphous, but it conserves a
memory of its original starting configuration. This memory
manifests upon pressure release: the sample recovers (partially)
the orientation it had prior to compression. We discuss the
characteristics of the interaction potential that are responsible
for the different behavior.

The paper is organized as follows. In Sec. \ref{s:model}, we present the model and
details of the performed  numerical simulations. The ideal martensitic
transformation is introduced in Sec. \ref{s:idealmarten}. In Sec. \ref{s:mt}, we
describe the martensitic transformation for the case when low energy barriers are
involved in the growth process. The case were this energy barriers are higher,
which leads to a WPIA, is discussed in Sec. \ref{s:QA}. Finally,
Sec. \ref{s:conclusions} contains some discussion and conclusions.

\section{The model, numerical details and equilibrium properties}\label{s:model}

\begin{figure}[!tbp]
\includegraphics[angle=-90,width=8cm,clip=true]{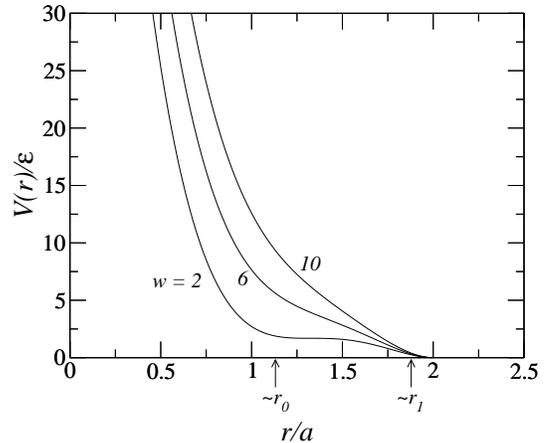}
\caption{\label{fpot} Pairwise core-softened interparticle
potential used in the present work [from Eq. (\ref{vder})]. The distances $r_0$ and
$r_1$ are schematically shown.}
\end{figure}

We consider a two-dimensional system of particles, interacting
through a specially devised isotropic, purely repulsive pairwise potential.
Its main characteristic is the existence of
two competing distances
$r_0$ and $r_1>r_0$ at which neighbor particles prefer to be
located, depending on the applied pressure.
This kind of
core-softened potential has been previously used to study
anomalous properties of tetrahedrally coordinated materials,
\cite{jagl98,jagl01,sadr98,scal01} and it was recently used to
study the TPIA case also.\cite{bust04} The interaction
potential $V(r)$ between two particles separated a distance $r$
is explicitly given by \cite{jagl01,bust04}
\begin{eqnarray}
\label{vder}
V(r)&=&\frac{\varepsilon}{0.81}[8(r/a-1.55)^2+16(r/a-1.55)^4+
\nonumber\\
& &w(r/a-2)^2+0.81]\text{     for } r<2a\\
&=&0\text{     for } r>2a  \nonumber
\end{eqnarray}
Here, $\varepsilon$ and $a$ fix the energy and length scales, and $w$ is a control
parameter that allows us to change the height of the shoulder of the interparticle potential
(see Fig. \ref{fpot}). Neighbor particles prefer to be located on both sides of the
shoulder, in the positions qualitatively indicated in Fig. \ref{fpot} as $r_0$ and $r_1$.
Simulation results will be presented mainly
for $w=6$ and $w=10$ to assess the effect of the barrier height
on the morphology of the microstructure obtained.
Note that, although for $w=10$ the curvature of the interparticle potential does
not change sign, the potential gives rise to two preferred distances when analyzing 
two- or three-dimensional configurations.

The system is simulated by standard molecular dynamics in the NVT
ensemble.\cite{allen,haile}
This choice allows us to survey all regions of the
volume-pressure curve, including those that would be unstable in
constant pressure simulations.
We perform simulations with
$9540$ particles, with a time step of $\delta t=0.01 t_0$ (where
$t_0=a\sqrt{m/\varepsilon}$ is the time unit,
$m$ being the mass of each particle). Pressure and elastic constant are calculated along the simulations using standard 
formulas (see the Appendix for details).
We always use periodic boundary conditions (PBC). This is a particularly convenient
choice when we want to avoid the heterogeneous nucleation of the new phase. On the contrary, in 
the case in which we study the heterogeneous nucleation, we introduce the nucleation centers by hand 
in the crystalline structure. The use of
PBC in this case does not impose any strong additional restriction in the possible transformations to be
observed, as our systems contain enough atoms to accommodate essentially any possible new phase, even using 
PBC.

The volume of the system is changed at a fixed rate
during the simulation.
This change is done by rescaling all coordinates of the particles and size of the
simulation box.
The volume change rate is taken to be as low as possible (within reasonable computational
time) to simulate a quasistatic process. However, it is of course a very large rate
in physical units. For instance, a typical volume change rate we use is
$10^{-4} t_0^{-1}$ , and for an atomic particle ($t_0 \sim 10^{-15}s$) this implies
a rate of the order of $10^{-2}$ per femtosecond.

By rescaling down the
velocities of the particles during the simulation, temperature is kept constant at
a very small value of the order of $10^{-5}\varepsilon$ (we take the
Boltzmann constant to be unity). As the transformation process
implies local mechanical instabilities, there is a systematic
transformation of potential energy into kinetic energy. In
experiments, this reflects an increase of the temperature of
the system. Here the kinetic energy excess is taken away
through the mentioned velocity rescaling procedure.

We emphasize that the use of a two-dimensional system is dictated
by the simplicity of the simulation and also of the interpretation of
the results. While we expect the same qualitative kind of
results to reappear in full three dimensional systems, the
computational time increases noticeably, and the adequate
identification of the phases becomes much more involved. Note that
in a previous work it has been shown that the number of
equilibrium  solid phases noticeably increases when passing from two to three 
dimensions,\cite{jagl99b} then adding to the complexity of
the three-dimensional case.

One of the most striking properties of particles interacting
through the potential described in the previous
section is that the stable crystalline ground state
is not necessarily the triangular lattice. In fact,
although at low (high) pressures the ground state of the system
is a triangular structure with a lattice parameter around $r_1$
($r_0$), at intermediate pressures lower energy configurations
can be obtained that have some of its first neighbors at
distances $\sim r_0$, and some others at $\sim r_1$. By
increasing the pressure, the configuration to which the low
density triangular (LDT) structure is expected to transform is
the chainlike (CH) structure.\cite{jagl98} This is the transition
that we will be mostly interested in, and we will analyze it
for two different values of $w$, namely $w=10$ and $w=6$. We will see that there
are noticeable differences between the two cases.

\section{The Ideal Martensitic transformation} \label{s:idealmarten}

\begin{figure}[!tbp]
\includegraphics[width=8cm,clip=true]{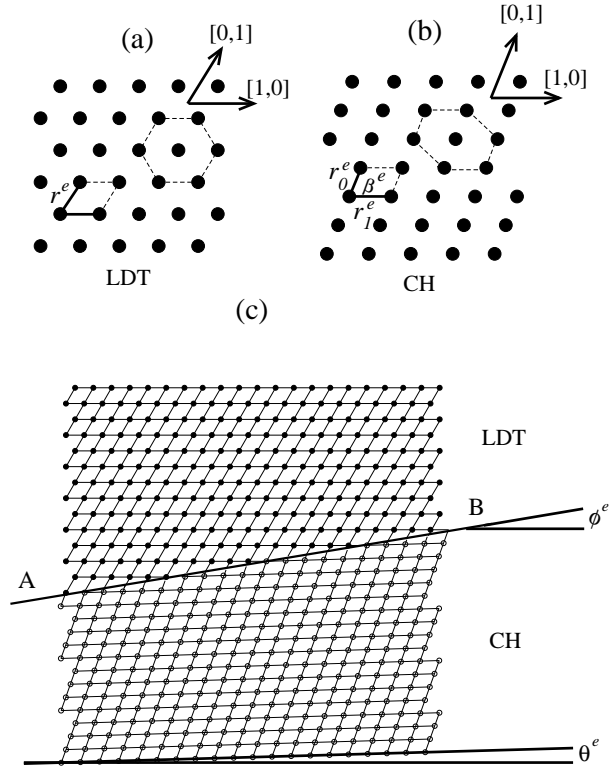}
\caption{\label{fdir} Ideal (a) LDT and (b) CH structures.
The lattice parameter and the crystallographic directions are shown, together
with the perfect and distorted hexagonal unit. (c) Interphase between the LDT and
CH structures at equilibrium. The invariant line AB is rotated $\phi^e$ with
respect to the LDT structure. The $[1,0]$ direction of the CH structure is
rotated $\theta^e$ with respect to the $[1,0]$ direction of the LDT structure.}
\end{figure}

We will study the behavior of the LDT phase upon volume reduction.
The LDT-CH transformation admits a natural description as a
martensitic transformation as depicted in Fig. \ref{fdir}. In
this description, the LDT corresponds to the austenite phase, and
the CH to the martensite phase. Upon volume reduction, the martensite phase
becomes stable above the
equilibrium transition pressure $P^e$. Two structures
can coexist at this pressure, namely a LDT structure with lattice
parameter $r^e$ [Fig. \ref{fdir}(a)], and a CH structure
(corresponding to an oblique two-dimensional Bravais lattice)
with parameters $r_0^e$, $r_1^e$, and $\beta^e$ [Fig.
\ref{fdir}(b)]. To minimize the interface energy at the equilibrium pressure, the two
structures have to match along a particular line, called the invariant line
or habit line (habit plane in three dimensions, line AB in Fig. \ref{fdir}).
This line is determined from the condition that
both phases are
unstrained away from it.\cite{nish78}
A relative rotation of the two phases is necessary in
order to define the invariant line.
The orientation of the invariant line can be easily
evaluated given the parameters of the martensite and the
austenite phase.  The structural parameters of the
austenite (LDT) and martensite (CH) phases at the equilibrium pressure
are given in Table \ref{t1}.

The ideal transformation between the LDT and CH structures does not imply bond
breaking, which means that each particle conserves the same first neighbors in the
two phases. This implies microscopic memory in the sense that each local
neighborhood transforms back to exactly the original configuration upon decompression.
Furthermore, this implies that after
a cyclic LDT-CH-LDT transformation, the original structure is recovered with the
same initial orientation. This property is directly related to the shape memory effect
of martensites.

If an invariant line separating the two coexisting phases has formed at the
transformation pressure, the transformation can proceed as volume is reduced
by reaccommodating  particles of the austenite close to the interface to fit into
the sites of the martensite.\cite{pond03}
The advance of the interface occurs through the lateral movement of the
steps that form the interface, and at the equilibrium pressure it
implies the surmounting of an energy barrier.
Then, at $T=0$ a small overpressure is necessary to climb this energy barrier.
As we will see, this ideal growth mechanism in the direction perpendicular to
the invariant line becomes more complicated when strains are involved, and
depends strongly on the height of the barrier for the transformation between
the austenite and martensite phases.

\begin{table}
\caption{\label{t1} Structural parameters at the LDT-CH
equilibrium point (see Fig. \ref{fdir} for definitions of the crystallographic parameters),
along with the pressure and volume values at the mechanical instability. The
angles $\theta^e$ and $\phi^e$ correspond to a relative rotation of the two
phases and a reference angle between the invariant line and the LDT
structure, respectively (see Fig. \ref{fdir}).}
\begin{ruledtabular}
\begin{tabular}{ccc}
Parameter&$w=6$&$w=10$\\
\hline
$P^e a^2/\varepsilon$&4.83&8.55\\
$v^e/a^2$(LDT)&3.01&2.74\\
$r^e/a$(LDT)&1.86&1.78\\
$v^e/a^2$(CH)&2.29&2.38\\
$r_0^e/a$(CH)&1.25&1.37\\
$r_1^e/a$(CH)&1.93&1.87\\
$\beta^e$(CH)&71.07&68.56\\
$\theta^e$(CH)&1.46&1.65\\
$\phi^e$(CH)&5.47&9.43\\
$P^c a^2/\varepsilon$&6.38&9.34\\
$v^c/a^2$&2.76&2.64\\
\end{tabular}
\end{ruledtabular}
\end{table}

\section{Nucleation and growth in the case of a low--energy barrier}\label{s:mt}

The previous ideal description of the martensitic transformation can differ
from the actual way in which
the martensite nucleates and grows
from the pure austenite phase in a real situation.
In this section, we study this process using the potential in Fig. \ref{fpot} with $w=10$.
On increasing pressure, the low
density triangular structure becomes thermodynamically unstable against the
chainlike structure beyond $P^e$.
However, as this transformation is first order, at $T=0$ the
perfect LDT structure survives in a
metastable state at higher pressures until it finally becomes mechanically unstable.
For the case of a perfect (without defects) starting sample, we call the instability pressure
$P^c$ (and the corresponding specific volume $v^c$).
The mechanical stability
limit corresponds to the value of pressure at which one of the normal modes
(phonons) of the original structure first destabilizes. The calculation of the
phonon structure of the triangular lattice is straightforward.
For our model potential Eq. (\ref{vder}) interactions with second and
further neighbors are identically zero, and the stability condition
can be written as
\begin{equation}
\label{stabcond} \frac{\partial^2 V}{\partial r^2} + \frac{3}{r}
\frac{\partial V}{\partial r} > 0,
\end{equation}
where $r$ is the nearest neighbor distance of the triangular
lattice. At the instability pressure $P^c$, three equivalent shear phonon
branches become zero energy. They correspond to shear phonons in
the three equivalent directions $[0,1]$, $[1,0]$, and $[-1,1]$
(Fig. \ref{fdir}). Note that, due to the lack of next nearest
neighbors interactions, the phonon branch along these directions
is of the type $\sim \sin(k)$, and at $P^c$ the whole phonon
branch becomes dispersionless.

The fact that the destabilization is initiated by a shear phonon
is consistent with an analysis in terms of Born inequalities
\cite{Born}. In fact, condition (\ref{stabcond}) is equivalent to
the generalization of Born stability criteria to stress dependent
conditions \cite{wang93,zhou96} in the static (zero temperature)
limit. In two dimensions, the Born stability criteria establish
that a system is stable if the stress dependent shear modulus $G'
= (C_{11}-C_{12})/2-P$ is positive \cite{zhou96}. For the zero
temperature case $G'=\sqrt{3}/4[V''(r)+3V'(r)/r]$
(see Appendix), and then the Born stability
criteria reduce to condition (\ref{stabcond}).

\begin{figure}[!tbp]
\includegraphics[angle=-90,width=8cm,clip=true]{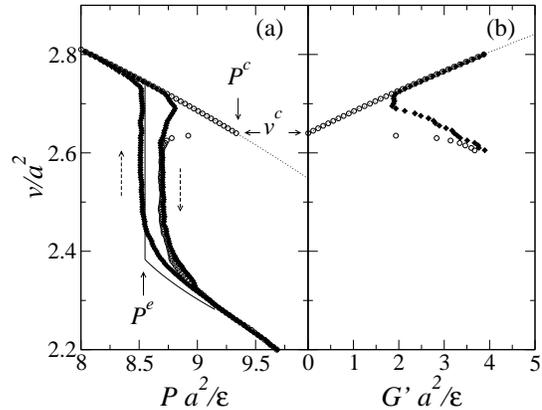}
\caption{\label{vp_w10} (a) Volume-pressure curves for $w=10$. The
dotted line is the analytic expression for the evolution of the LDT
structure if this remains stable, and the thin continuous line marks the equilibrium
transition to the CH structure and the subsequent evolution
of this structure. The full and open symbols correspond to
simulations with and without a vacancy in the original sample.
Vertical dashed arrows indicate the compression-decompression paths.
(b) Evolution of the stress dependent shear modulus
$G'=(C_{11}-C_{12})/2-P$ upon decompression for $w=10$.
The dotted line is the analytical result for $G'$ evaluated for the perfect triangular lattice
(see the Appendix).
Note how in the defect free case, the approach to the transition point ($v\rightarrow v^c$) is
signaled by the vanishing of the stress dependent shear modulus.}
\end{figure}

The results of numerical simulation [Fig. \ref{vp_w10}(a)] confirm
the previous analysis. Starting from a perfect triangular
structure at zero temperature, once the system reaches the
instability volume $v^c$, the triangular configuration
destabilizes
and a new stable configuration should emerge. The
evolution of the stress dependent shear modulus
upon volume
decrease is shown in Fig. \ref{vp_w10}(b), where it is clear that
it vanishes at the mechanical instability point.

The consideration of a perfect sample without defects is
important in order to realize the equivalence of the mechanical
instability limit with the vanishing of the energy of a phonon
mode. Moreover, in the present case it also coincides with the
vanishing of an elastic constant (a long wave length elastic
constant), because of the lack of interactions to next nearest
neighbors.\cite{nota1} But certainly, a defect free sample is not typically
encountered in experiments. It is important to realize that even
tiny amounts of defects can produce important changes in the
triggering of the transformation. For instance, a single vacancy
can destabilize the lattice at pressures much lower than $P^c$.
This was shown already in the case of TPIA,\cite{bust04} and it is also true
here. The evolution of pressure upon volume decrease for a lattice with a single
vacancy is also shown in
Fig. \ref{vp_w10}(a).
We see that a single vacancy is able to destabilize the
original triangular lattice without essentially any pressure overshooting.
Note, however, that the inclusion of a single vacancy has no
appreciable effect on the values of the shear modulus prior to
the transition [Fig. \ref{vp_w10}(b)].
This means that even in the
presence of tiny amounts of disorder, the examination of the
elastic constants (or even of the whole phonon spectrum) will not
indicate the approach to the instability point. In fact, the
instability occurs because of the destabilization of localized
modes close to the defect, which have no effect on the spectrum
of the (extended) phonons. The absence of macroscopic effects
signaling the approach to the instability is usually associated
to thermally activated first order transitions. 
We see that a similar phenomenology occurs here in the absence of
thermal fluctuations, and within a scenario of a transition
driven only by mechanical instabilities.

\begin{figure}[!tbp]
\includegraphics[width=8cm,clip=true]{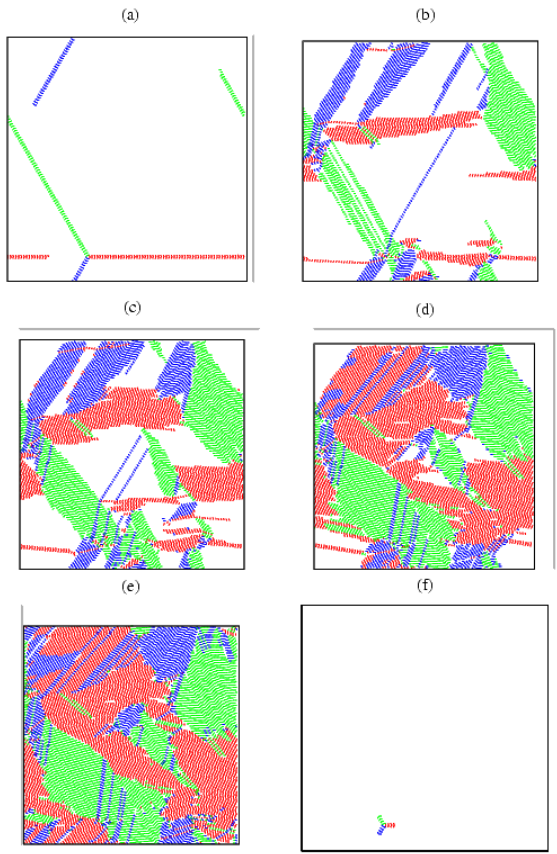}
\caption{\label{snap_w10}(Color online) Snapshots of the system with
$w=10$ corresponding to: (a) $v/a^2=2.7$, (b) $v/a^2=2.6$, (c)
$v/a^2=2.5$, (d) $v/a^2=2.4$, (e) $v/a^2=2.3$, on compression, and (f)
$v/a^2=2.8$ after decompression from the configuration in (e).
Only collapsed particles are shown joined by segments, with three different colors to identify
the three different variants of the martensite (intact austenite
is not plotted for clarity). The starting system contains a single
vacancy at the position where the three different variants start
nucleating in (a). Note that periodic boundary conditions are used.}
\end{figure}

In Fig. \ref{snap_w10}, a sequence of snapshots is shown to follow the
morphology of the transformed regions in the case in which a
single vacancy is present in a monocrystalline sample. We can
see that the first instability produces the collapse of adjacent
rows of particles, around the defect. These collapses generate
three platelets that propagate out of the vacancy [Fig. \ref{snap_w10}(a)]. The collapse
directions in the three platelets are different, and are related
to the three different variants of the martensitic
transformation. In this stage, the
martensite crystallites are compressed in the longitudinal
direction, in order to match the lattice parameter of the
original triangular structure.\cite{pond03} In a subsequent stage [Fig. \ref{snap_w10}(b),(c)], the new
crystallites reaccommodate to take the form of wedge shaped crystallites, with the
interface between the martensite and austenite phase being very close
to the ideal habit line of the transformation. Then upon further compression
[Fig. \ref{snap_w10}(c-e)] the crystallites grow perpendicularly to the habit line,
very much as described for the ideal case in the previous Section.
The final state of the system (when the whole sample has
transformed to the martensitic phase) consists of a
polycrystalline sample with the three different orientation of
the possible variants [Fig. \ref{snap_w10}(e)]. Note from Fig. \ref{vp_w10}(a),
that after nucleation, pressure has
remained essentially constant during the whole transformation from
the LDT to the CH structure, at a value slightly larger than the equilibrium value.
This value reflects the overpressure needed to
keep the transformation proceeding.

\begin{figure}[!tbp]
\includegraphics[angle=-90,width=8cm,clip=true]{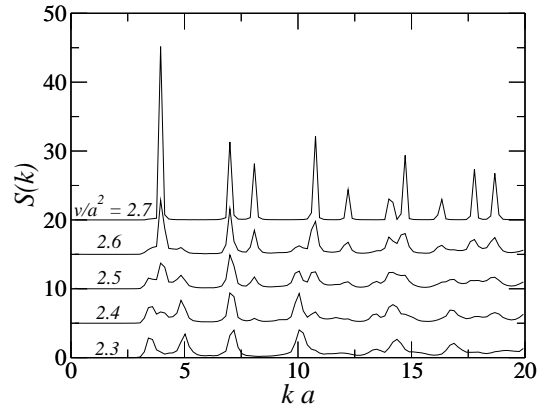}
\caption{\label{vsdk_w10} Evolution of the averaged structure factor $S(k)$ with
decreasing volume for the LDT-CH transformation with $w=10$ (the curves are vertically displaced,
for clarity). Different curves
correspond to the configurations in Fig. \ref{snap_w10} (a)-(e).}
\end{figure}

\begin{figure}[!tbp]
\includegraphics[width=8cm,clip=true]{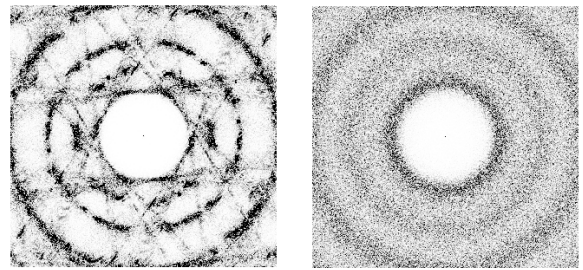}
\caption{\label{sdk_w10}
Two-dimensional (upper) and angularly averaged (lower) structure factors of the configuration in Fig. \ref{snap_w10}(e), and of the configuration obtained by quenching from the melt at the same density.
In the top, the two-dimensional structure factors of the polycrystalline CH structure (left) and quenched amorphous structure (right) are shown.
In the lower figure, bold and thin lines correspond to the averaged $S(k)$ for the polycrystalline CH and quenched amorphous structures, respectively.
There is a clear crystalinity observable in the first case, even if we look only at the
radial structure function. This sample should be characterized as polycrystalline.}
\end{figure}

If the volume of the system is increased back starting from the
sample completely in the martensitic phase, upon complete
pressure release essentially any particle in the system returns
to its original position, in which it has exactly the same
neighbors as before compression. In particular, this implies that
the original monocrystalline austenite sample is reobtained, with
the same orientation as in the starting configuration.\cite{nota2}
This is not a surprise, since it is in the essence of the shape
memory effect of martensites. It can be seen in Fig.
\ref{vp_w10}(a) that the return occurs with a noticeable
hysteresis with respect to compression. This hysteresis is a
consequence of the individual hysteresis of pairs of neighbor
particles on passing between the two equilibrium distances, and
can be used to quantify the energy dissipated during a
compression-decompression cycle.

Now we want to discuss how the transformation manifests when we
observe the diffraction patterns of the structures. In Fig.
\ref{vsdk_w10} we show the angularly averaged structure factor
$S(k)$ corresponding to the configurations in Fig.
\ref{snap_w10}(a)-(e). It is possible to see how some
characteristic peaks of the LDT structure progressively disappear
while new peaks appear that correspond to the CH structure. A
broadening of the peaks is also apparent, due to the finite size
of the crystallites in the new configuration. This broadening
however is not large enough to consider the sample as X-ray amorphous.
By comparing the structure factor of this structure with that of
a sample obtained by quenching from the melt at the same density
(Fig. \ref{sdk_w10}), we first observe that the two-dimensional
structure factor displays clear evidence of the polycrystalline
character of the sample. But even if we look only at the radial
structure factor, there are much sharper peaks for our sample
compared to the quenched liquid. Then the configuration in Fig.
\ref{snap_w10}(e) can be correctly identified from its structure
factor as a polycrystalline sample.

\section{Blocked growth for large energy barrier}\label{s:QA}

\begin{figure}[!tbp]
\includegraphics[angle=-90,width=8cm,clip=true]{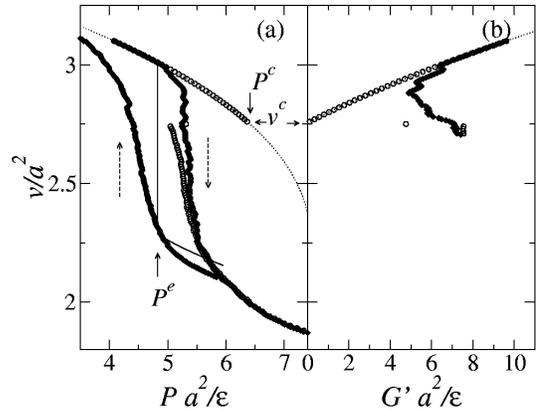}
\caption{\label{vp_w6} (a) Volume-pressure curves for $w=6$. The dotted line
is the expected evolution of the LDT structure, and the thin continuous line marks
the equilibrium transformation to the CH structure and the subsequent evolution
of the new structure. The full and open symbols correspond to simulations
with and without a vacancy in the original sample.
Vertical dashed arrows indicate the compression-decompression paths.
(b) Evolution of the stress dependent shear modulus
$G'=(C_{11}-C_{12})/2-P$ upon decompression for $w=6$.
The dotted line corresponds to $G'$ evaluated from the
interatomic potential (see the Appendix).
$P^c$ and $v^c$ are given by the vanishing of the stress dependent shear modulus in (b) (see Table \ref{t1}).}
\end{figure}

\begin{figure}[!tbp]
\includegraphics[width=8cm,clip=true]{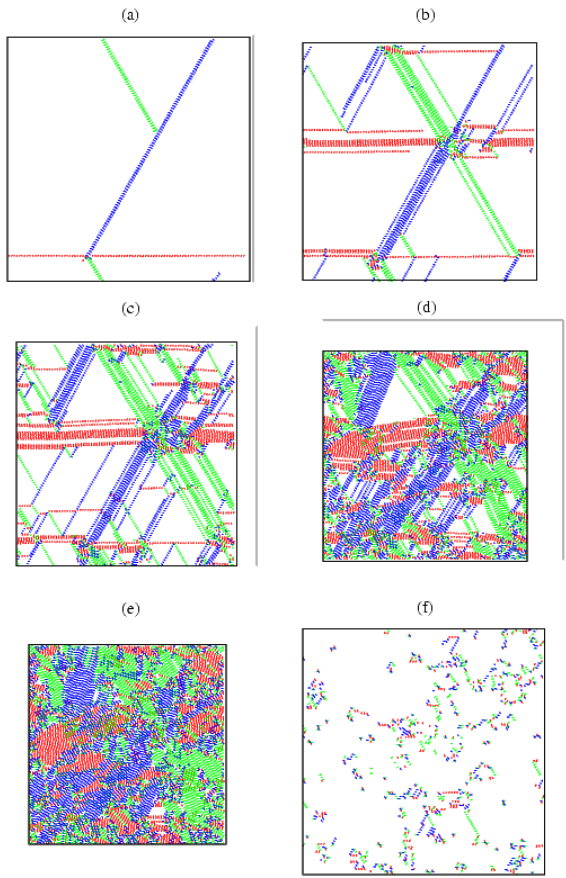}
\caption{\label{snap_w6} (Color online) Snapshots of the volume controlled simulation of the system with
$w=6$ corresponding to: (a) $v/a^2=2.95$, (b) $v/a^2=2.8$, (c)
$v/a^2=2.6$, (d) $v/a^2=2.3$, (e) $v/a^2=2.1$, on compression, and (f)
$v/a^2=3.05$ on decompression from $v/a^2=2.1$.
Only collapsed particles are shown joined by segments, with three different colors to identify
the three different variants of the martensite (intact austenite
is not plotted for clarity). The starting system contains a single
vacancy at the position where the three different variants start
nucleating in (a). Note that periodic boundary conditions are used.}
\end{figure}

In the present section, we will study the modification brought upon by a change of the parameter $w$ of the
potential. The value used here is $w=6$. In principle, the same description of the ideal martensitic
transformation is valid also in this case. The value of the crystallographic parameters are
given in Table \ref{t1}.
In Fig. \ref{vp_w6}(a), we show the volume-pressure evolution of the volume
controlled compression for $w=6$. Open symbols correspond to the starting
sample without defects. The simulation shows that the system remains in the triangular structure
until the mechanical
stability limit $(P^c,v^c)$ is reached. On the contrary, the
sample with a single vacancy begins to transform when
local mechanical instabilities occur, before reaching the mechanical
stability limit. The behavior of the stress dependent shear modulus for the
present case, with and without the vacancy, is shown in Fig. \ref{vp_w6}(b).
The phenomenology up to here is very similar to the case of the previous
section. However, important differences appear when we look at the morphology
of the samples and the structure factors.
We again focus on the case of a system with a single vacancy.
In Fig. \ref{snap_w6}, snapshots of the system as volume is decreased are shown.
As in the previous case, the first elements of the new phase to be
nucleated are the platelets starting at the position of the vacancy.
But contrary to the previous case, they do not rearrange to generate an interface
along a habit line. As the growth process in these conditions implies much higher
overpressure, the growth remains mostly
blocked, and upon further compression nucleation of new platelets
throughout the sample occurs. New and old platelets interact elastically, and deform each other,
generating at the end a structure
that in spite of being martensitically related
to the original one, contains crystallites that are extremely small.

On very general grounds, since
our polycrystalline structure has very small crystallites,
a broadening of the characteristic peaks of the azimuthally averaged X-ray
pattern is expected. \cite{yama97} There is a continuous crossover however from this
`broadened peaks' scenario, to one in which
no sharp peaks are observed. In the last case the
sample should be characterized as X-ray amorphous. But even in this case it can
possess specific crystal
characteristics, and in particular it may conserve a structural relation with the
parent structure that will manifest upon decompression.

\begin{figure}[!tbp]
\includegraphics[angle=-90,width=8cm,clip=true]{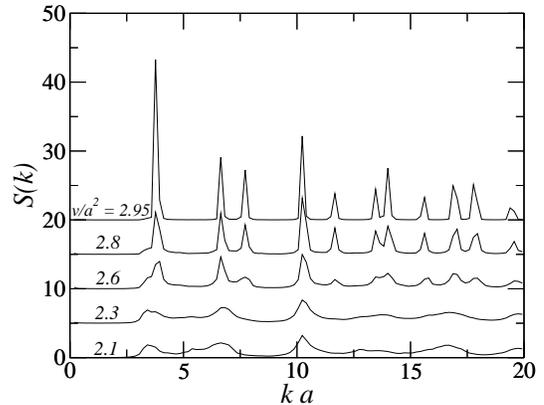}
\caption{\label{vsdk_w6} Evolution of the averaged structure factor $S(k)$ with
decreasing volume for the LDT-CH transformation with $w=6$. Each curves
correspond to the configurations in Fig. \ref{snap_w6} (a)-(e).}
\end{figure}

\begin{figure}[!tbp]
\includegraphics[width=8cm,clip=true]{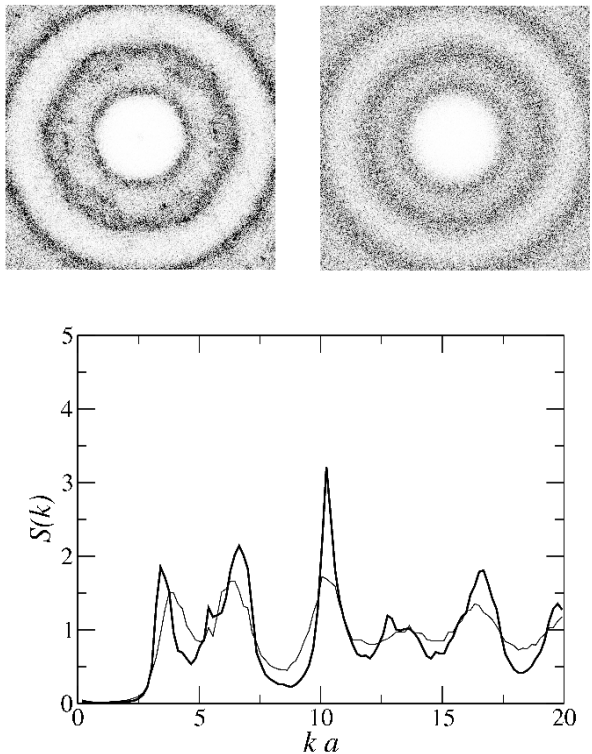}
\caption{\label{sdk_w6} Comparison of the structure factor corresponding to the
sample obtained in the LDT-CH transformation [Fig. \ref{snap_w6}(e)] with a
quenched amorphous structure at the same density.
In the top the two-dimensional structure factors for the polycrystalline
CH structure (left) and quenched amorphous structure (right) are shown. The angular
averaged $S(k)$ is plotted at the bottom for the polycrystalline CH structure
(bold continuous line) and quenched amorphous structure (thin continuous line).}
\end{figure}

In Fig. \ref{vsdk_w6}, we show angularly averaged structure factors corresponding to
the configurations shown in Fig. \ref{snap_w6} (a)-(e). Again, the peaks of the
LDT structure disappear while reducing the volume, but the intensity of
the peaks corresponding to the CH structure is much weaker than in the $w=10$ case.
In Fig. \ref{sdk_w6}
the two-dimensional X-ray diffraction patterns
of the structure obtained from a single crystal LDT sample, and
an amorphous structure obtained by quenching from the melt (with
the same density), are shown.
We see that the two-dimensional pattern of the sample obtained by
compression possesses an angular structure reminiscent of the
original crystal orientation (in fact, this structure will be
responsible for a `memory effect' when the sample is
decompressed, see below). However,
if our starting sample is already polycrystalline, as is usually the case experimentally, the X-ray
pattern is by definition isotropic. In that case all the
information available is the radial X-ray pattern. The angularly averaged $S(k)$
are shown at the bottom of Fig. \ref{sdk_w6}.
On the basis of the $S(k)$ function, the
compressed sample cannot be clearly distinguished from that obtained by
quenching from the melt. This sample is then X-ray amorphous.
Thus, we have here a sample that has WPIA.

\begin{figure}[!tbp]
\includegraphics[angle=-90,width=8cm,clip=true]{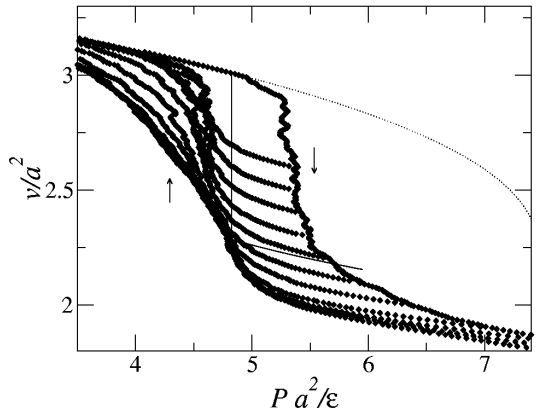}
\caption{\label{mvp_w6}$v-P$ compression-decompression cycles for
different values of the final volume reached. From bottom to top
$v_{min}/a^2$ goes from $1.6$ to $2.6$, in steps of $0.1$. Dotted
and thin continuous lines as in Fig. \ref{vp_w6}.}
\end{figure}

The fact that this sample is X-ray amorphous, but still has
within its structure some information of the parent crystalline
phase from which it was obtained may be related to the so-called
memory glass effect (MGE).
The MGE was first identified in
AlPO$_4$ berlinite,\cite{krue90,tse92} and then it was recognized
to occur in a variety of materials such as SnI$_4$ and
LiKSO$_4$.\cite{shar96} The effect consists in the amorphization
of a sample under pressure, and the successive recovery of a
crystalline sample, with the \textit{same original orientation} upon
decompression.
In the following, some characteristics of the MGE for the present model
system are studied.

\begin{figure*}[!tbp]
\includegraphics[width=1\linewidth,clip=true]{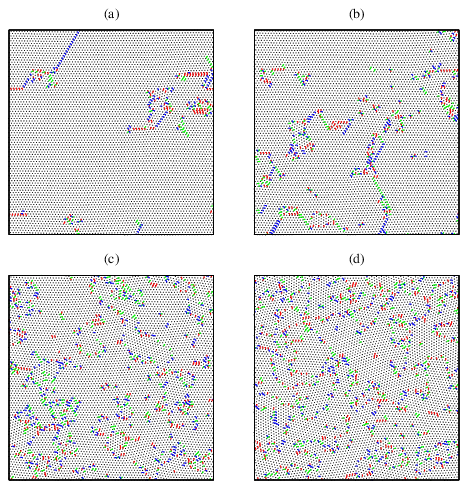}
\caption{\label{snap_m} (Color online) Snapshots of the system as recovered from the
compression-decompression cycles shown in Fig. \ref{mvp_w6}. All the snapshots
correspond to a final volume of $v/a^2=3.05$. The minimum volumes reached are (a)
$v/a^2=2.4$, (b) $v/a^2=2.2$, (c) $v/a^2=2.0$ and (d) $v/a^2=1.8$.
Different colors indicate particles collapsed along each of the directions of the possible martensitic variants.
Dots corresponding to the position of the particles are shown in order to
identify the appearance of polycrystallinity.}
\end{figure*}

The complete $v-P$ curves during volume controlled
compression-decompression cycles are shown in Fig. \ref{mvp_w6}
for different values of the final volume reached, $v_{min}$.
The final configurations after the cycling are shown in Fig. \ref{snap_m}. All
the snapshots correspond to a recovered volume of $v/a^2=3.05$.
Shown are the defects introduced by the cycling, corresponding to small regions
of the different martensitic variants.
Small dots correspond to the position of the particles.
On a direct examination of the recovered structures, it is seen that the system
has a tendency to recuperate its original orientation upon decompression.
A closer inspection
shows that the recovered LDT structure retains its initial orientation for
high $v_{min}$, but becomes polycrystalline for lower $v_{min}$, losing the
`memory' of the initial orientation.

\begin{figure}[!tbp]
\includegraphics[angle=-90,width=8cm,clip=true]{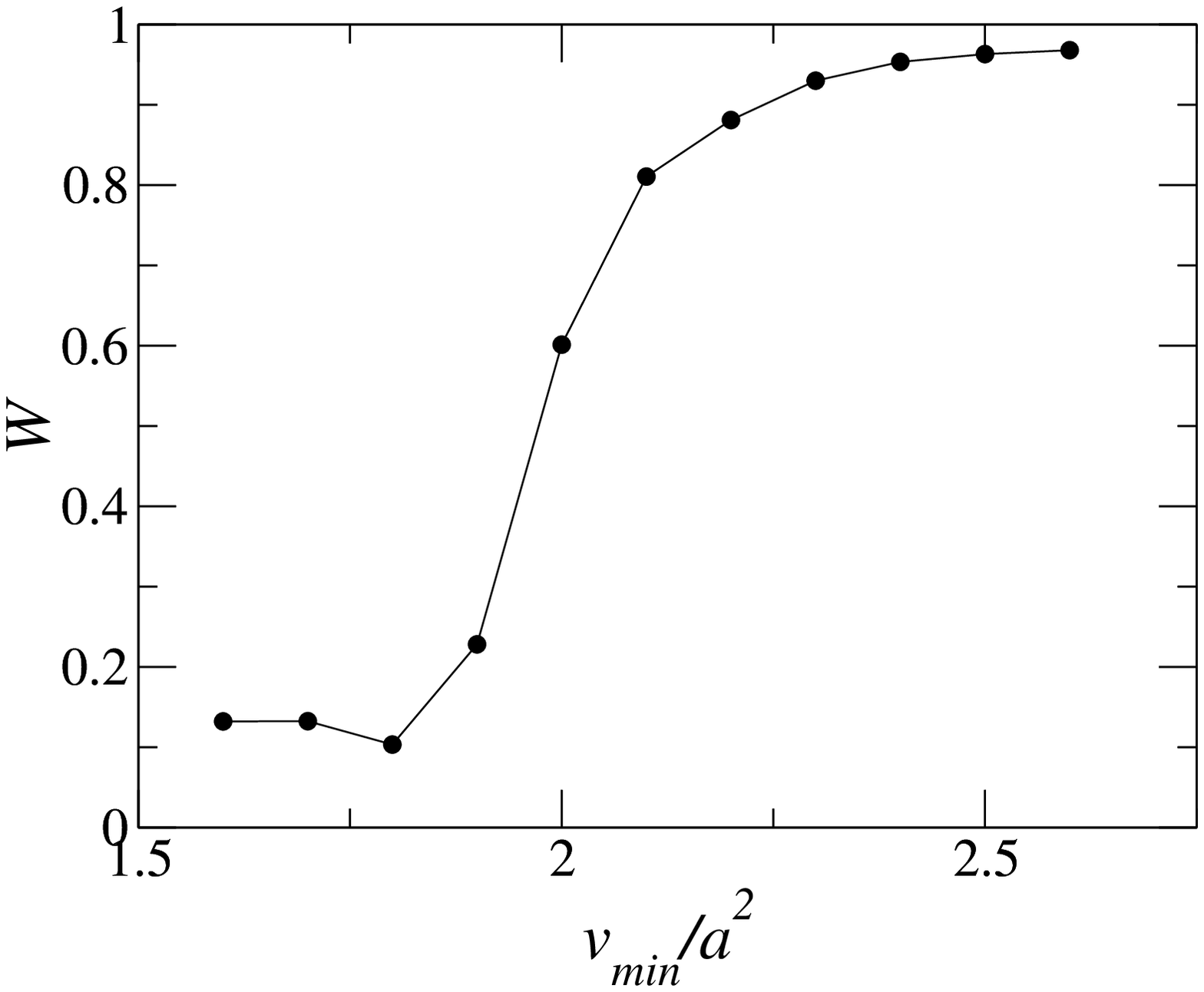}
\caption{\label{mw} The parameter $W$ defined in Eq. (\ref{wdef}) as a function
of the minimum volume $v_{min}$ reached in the compression-decompression cycles
shown in Fig. \ref{mvp_w6}.}
\end{figure}

It is convenient to have a single parameter that measures
the degree of memory of the structure.
We take this parameter to be $W$, defined as
\begin{equation}
\label{wdef}
W=\frac{1}{36N}\sum_j \text{Re}\left[\sum_{k,k'}
e^{6i(\alpha_{jk}-\alpha_{jk'})}\right],
\end{equation}
where $k$ and $k'$ are the nearest neighbors of particle $j$ before and
after the compression-decompression cycle, and $\alpha_{jk(k')}$ is the angle
between $\vec{r}_{k(k')}-\vec{r}_j$ and an arbitrary reference direction.
In particular, it can be seen that if the sample returns exactly to its initial
configuration, we get $W=1$. On the other hand, if the final local orientations
in the sample are completely uncorrelated with the original ones, then $W=0$.
We take as the reference volume to compute $W$ the value $v/a^2=3.05$ and
the plot of $W$ as a function of the minimum volume $v_{min}$ reached during
compression is seen in Fig. \ref{mw}.
We see in fact, that this number is close to one for $v_{min}/a^2>2$, whereas
it is essentially zero when $v_{min}/a^2<1.9$.
The memory of the system is lost continuously when $v_{min}$ becomes lower,
but note the interesting fact that for $v_{min}=2.1$, in which there is no
LDT phase left in the compressed sample (see Fig. \ref{snap_w6}), the value
of $W$ is still clearly different from zero.
All the memory of the system here can be associated to the preferred
orientation of the crystals of the CH phase, which in turn occurs because of
the martensitic nature of the LDT-CH transition.
As for this volume the sample is X-ray amorphous, this is an example of MGE.

\section{Discussion and conclusions}\label{s:conclusions}

We have studied a two dimensional model at $T=0$ that displays a pressure driven
martensitic transformation between a triangular structure
(the austenite) with
rotational $C_6$ symmetry, and a chain like structure (the martensite) with a lower
($C_2$) rotational symmetry. The martensite phase
can be considered to be obtained from the austenite
through a compression in one direction and an expansion in the perpendicular direction.
The transformation occurs with an appreciable volume change. This is natural since pressure (which couples
to volume in the free energy) is the driving force of the transformation.

The transition does not spontaneously occur at the equilibrium transformation
pressure because an energy barrier exists. An excess pressure is needed to
trigger the transformation. Defects in the original crystalline
structure produce the appearance of `reaction paths' which have
(at the equilibrium pressure) relatively low energy barriers.
These barriers vanish with a relatively low overpressure,
triggering the transition. The necessary overpressure in the
presence of defects is typically much smaller than that
needed to destabilize the defect free structure.
We have shown in particular, that vacancies are very effective
in favoring the nucleation of three platelets related to the three
variants of the martensite, which are however elastically distorted with
respect to the ideal martensite. We call them pseudo-martensitic
crystals.\cite{pond03}

After the nucleation of pseudo-martensite platelets, two different evolutions were observed.
In one case, the pseudo-martensitic
platelets find the way to become wedge shaped crystallites of the
true martensitic structure. In this process, the internal stress of the
martensite relaxes, and the interface between austenite and martensite
becomes very nearly the ideal invariant line of the transformation. These
wedge shaped crystals grow upon volume decrease, until the whole sample
consists of a collection of crystallites of the three different variants
of the martensite, and no austenite remains.
In this case, the final size of the crystallites is expected to depend mainly on the density
of defects in the system  and the compression rate, since there are no other important limitations
to martensite growth. In principle, very large crystal sizes can be obtained in samples with
low concentration of defects driven at very a low compression rates.

In a second case, the conversion of pseudo-martensite to true martensite is
strongly inhibited. In this case, upon compression, new platelets of
pseudo-martensitic structure nucleate,
that interact with preexisting ones. The structure obtained when the
austenite has completely disappeared is much more disordered than in the
previous case, and in particular, the final crystallite size in this case
is expected to remain of the order of few interatomic distances. In some cases, the
angularly averaged diffraction pattern of the final structure
may well be taken  by that of an
amorphous structure. We have called this
phenomenon weak pressure induced amorphization. This is not
a true pressure induced amorphization since the final structure has a
`memory' of its parent crystal that is observable, for instance, in the full
two-dimensional diffraction pattern. Upon decompression of this
sample, a defective crystal is obtained, which, however,
has a consistent tendency to be oriented in the same form as the original
parent crystal. This is a description of the
`memory glass effect' (MGE), observed in different experimental situations.

The two different behaviors were obtained by changing the parameters of the
interparticle potential.
Concretely, the easy growth of the martensite was observed when the barrier
between the two possible interparticle distances was small, whereas inhibited
growth was observed when this barrier was higher.

Although the present model is not aiming to reproduce the characteristics of a specific
real system with its complex interactions, some similarities concerning PIA,
martensitic transformation and the MGE, may be addressed between this
simple model system and real systems.

As we neglect temperature effects in our molecular dynamics
simulation, the transformation between the austenite and martensite
phases is of a mechanical nature (and this should be qualitatively correct for any
$T<T_g$). This means that the austenite is
driven into a metastable region, until a mechanical instability limit is
reached, and the martensite starts to nucleate spontaneously. Note the important fact (already present in our
previous results in \cite{bust04}) that once nucleated, the interface between the two phases 
can propagate spontaneously, and this propagation, which occurs essentially at the sound
velocity, is not limited by  the diffusivity of the particles in any of the two bulk phases
(which is zero here).
In the case of defect free samples the mechanical instability of the austenite
is signaled by the softening
of a phonon mode. However, in
the presence of defects the first mechanical instability
corresponds to localized modes around the defects, and therefore
it is not observable in the phonon spectrum. Thus, our present
and previous \cite{bust04} results reinforce our view that PIA is
a mechanical transition, i.e. always triggered by
mechanical instabilities. 
We find that defects can cause the transition pressure to be depressed to a value
very close to the ideal thermodynamical equilibrium pressure between the two phases.
We recognize however, that this result can be not universal but system dependent.
The relevance of defects density in
lowering the amorphization pressure was recently experimentally
highlighted, \cite{joha04} although whether the amorphization pressure
approaches the extrapolated melting line upon introduction of more defects is not clear. 
The experimental evidence of 
amorphization occurring near the extrapolated melting line in some systems is at the base
of a `thermodynamical' scenario for PIA
as directly related to the  `thermodynamical melting' mechanism at higher temperatures.
We see that this evidence is also fully compatible with a mechanical mechanism.


A transformation reminiscent of our martensitic transformation with easy growth
occurs between the high pressure phases of GaPO$_4$. In this case, the
low-cristobalite phase of GaPO$_4$ transforms to a \textit{Cmcm} structure on
pressure loading.\cite{robe94,shar95} The transformation
is of a displacive nature, and the mechanism consists of a fourfold-to-sixfold
coordination change of the gallium coordination shell.\cite{robe94}
The peaks of the polycrystal are clearly identified, which indicates a large crystal size.
In view of the results presented here, it could 
be interesting to exploit the martensitic point of view in analyzing this kind of pressure induced transformation.

On the other hand, the transformation in the case of blocked growth is in line with
investigations of the high pressure transformations of AlPO$_4$ and the MGE.
\cite{king94,garg00,shar00,gill95,rama03}
What was first viewed as a true PIA with MGE, was later identified as a
crystal-crystal pressure-induced transformation between a berlinite AlPO$_4$ structure and
a \textit{Cmcm} polycrystalline structure with very small crystal size,
i.e., in our language, a WPIA.
Carefully done molecular dynamics simulations \cite{garg00} and X-ray
experiments \cite{shar00} have identified some peaks of the X-ray diffraction
patterns of the high pressure phase as corresponding to the \textit{Cmcm} structure.
In addition, recent work has been devoted to establish
that the \textit{Cmcm} structure is the stable one at high pressures,
\cite{rama03} but little is known about the mechanism of the
transformation.
In view of the similarities between the MGE in our two-dimensional model and in the
AlPO$_4$ case, we suggest that MGE occurs simply as a consequence
of the amorphization being only of the weak type, instead of a
true one.
Furthermore, we speculate that the mechanism of the berlinite-\textit{Cmcm}
transformation in AlPO$_4$ could be very
similar to the mechanism of the low cristobalite-\textit{Cmcm}
transformation in GaPO$_4$. However, the growth of the  \textit{Cmcm}
structure over the berlinite phase of AlPO$_4$ could be inhibited by
energetic barriers, leading to the polycrystal with very small crystal size.

To conclude, we notice how the present model, although highly simplified and not representing
any particular real system, captures much of the phenomenology observed in a 
variety of real cases. As such, it is an important theoretical tool 
to guide the analysis of concrete cases. We expect to be able to go further in this direction in the future.

\begin{acknowledgments}

S.B. is financially supported by CONICET (Argentina), and also acknowledges
the hospitality of The Abdus Salam ICTP (Trieste), where part of this
work was done.

\end{acknowledgments}

\appendix*

\section{Stress dependent elastic constants and their static limit}

Fluctuation formulas for stress-free elastic constants in the NVT ensemble are given by\cite{zhou96}
\begin{eqnarray}
\label{Cijkl}
C_{ijkl}&=&\frac{2NkT}{V} \left( \delta_{ij}\delta_{jk} + \delta_{ik}\delta_{jl}
\right) - \nonumber \\
& &- \left( \langle P_{ij} P_{kl} \rangle -
\langle P_{ij}\rangle \langle P_{kl}\rangle \right) + B_{ijkl},
\end{eqnarray}
where $N$ is the number of particles, $k$ is the Boltzmann constant, $T$ is the
temperature and $V$ is the volume of the system. The brackets, $\langle...\rangle$, denotes
ensemble averages. $P_{ij}$ are the components of the
pressure tensor given by
\begin{equation}
\label{Pij}
P_{ij}=\frac{1}{V}\left( \sum_{\alpha}
\frac{p_{(i)\alpha} p_{(j)\alpha}}{m_{\alpha}} -
\sum_{\alpha < \beta} \frac{V'(r_{\alpha \beta})}{r_{\alpha \beta}}
r_{(i)\alpha \beta} r_{(j)\alpha \beta} \right),
\end{equation}
with $\alpha$ and $\beta$ particle running indexes, $p_{(i)\alpha}$ the momentum
of the particle $\alpha$ (with mass $m_{\alpha}$)  in the $i$ direction, and
$r_{\alpha \beta}$ the distance between particles, with component
$r_{(i)\alpha \beta}$ in the $i$ direction. Finally,
\begin{eqnarray}
\label{Bijkl}
B_{ijkl}&=&\frac{1}{V} \langle \sum_{\alpha < \beta} \left(
\frac{V''(r_{\alpha \beta})}{r_{\alpha \beta}^2}-
\frac{V'(r_{\alpha \beta})}{r_{\alpha \beta}^3} \right) \times \nonumber \\
& &\times r_{(i)\alpha \beta} r_{(j)\alpha \beta} r_{(k)\alpha \beta} r_{(l)\alpha \beta}
\rangle.
\end{eqnarray}
Here, $V'(r)$ and $V''(r)$ correspond to first and second derivatives of the
interaction potential energy at $r$.

Consider the static (zero temperature) limit. In this limit one has to consider
only the perfect crystal structure without any thermal fluctuation.
In this case the first term of Eq. (\ref{Cijkl}), proportional to the temperature, vanishes. The second term, which measures the fluctuation of the pressure tensor, vanishes too.
Besides,  the term containing the momentum
of the particles in Eq. (\ref{Pij}) is zero, and $C_{ijkl}=B_{ijkl}$. Then, in this limit,
the calculation of the elastic constant is straightforward, and instead of using the complete fluctuation formulas, which needs an
ensemble average to be performed, we use the static limit Eq. (\ref{Bijkl}) calculated over a given sample.

Let us consider, for instance, the calculation of $C_{12}$ in a triangular structure of lattice
parameter $r$ with nearest neighbors interactions only.
\begin{eqnarray}
\label{C12}
C_{12}&=&\frac{1}{2V}\sum_{\alpha,\beta=1}^N
\left(V''(r_{\alpha \beta})-\frac{V'(r_{\alpha \beta})}{r_{\alpha \beta}}\right)
\frac{r_{(x)\alpha \beta}^2 r_{(y)\alpha \beta}^2}{r_{\alpha \beta}^2}
\nonumber \\
&=& \frac{1}{2V} \left(V''(r)-\frac{V'(r)}{r}\right) \frac{1}{r^2}
\sum_{\alpha=1}^N 4 \left(\frac{r}{2}\right)^2
\left(\frac{\sqrt{3}}{2}r\right)^2 \nonumber \\
&=& \frac{1}{2V} \left(V''(r)-\frac{V'(r)}{r}\right) N \frac{3}{4} r^4
\nonumber \\
&=& \frac{\sqrt{3}}{4} \left(V''(r)-\frac{V'(r)}{r}\right),
\end{eqnarray}
where we used $N/V=2/(\sqrt{3}r^2)$. Analogously we obtained
\begin{equation}
\label{C11}
C_{11}=\frac{3}{4}\sqrt{3}\left(V''(r)-\frac{V'(r)}{r}\right),
\end{equation}
and
\begin{equation}
\label{P11P22}
P=-\sqrt{3}\frac{V'(r)}{r}.
\end{equation}
The stress dependent shear modulus is given by
\begin{equation}
\label{GP}
G'=\frac{C_{11}-C_{12}}{2}-P=\frac{\sqrt{3}}{4}
\left(V''(r)+3\frac{V'(r)}{r}\right).
\end{equation}
Then, the condition of vanishing stress dependent shear modulus coincides
with condition (\ref{stabcond}) in the static limit.

\bibliography{marten}

\begin{thebibliography}{46}
\bibitem[{not({\natexlab{a}})}]{nota1}
\bibinfo{note}{Next nearest neighbors interactions may eventually produce the
  instability to appear at some finite wave vector. If this is the case, then
  the instability will not be seen in the elastic constants, which are the long
  wavevector limit of the phonon spectrum.}
\bibitem[{not({\natexlab{b}})}]{nota2}
\bibinfo{note}{Note that this situation is much more restrictive than what is
  usually called in the literature a `reversible amorphization' transition. A
  reversible amorphization is usually meant to occur when a system passes from
  crystalline to amorphous upon compression, and becomes crystalline again upon
  decompression, without requiring any kind of `memory' between the initial and
  final crystal states. In this sense, all the results presented here show
  `reversible' amorphization. It has to be mentioned that we can obtain with
  our model `irreversible' amorphization (in which the sample remains amorphous
  when pressures is released) by the inclusion in the interaction potential of
  an attractive term among the particles, as discussed in Ref. [22]. To avoid
  confusion, we do not use the term `reversible' in the text.}
\end{thebibliography}
\end{document}